\documentclass[12pt]{article}
\usepackage{amssymb,amsmath}
\usepackage[pdftex]{graphicx}
\usepackage{subfig}
\usepackage{color}
\usepackage{float}
\usepackage{amsmath}
\usepackage{graphicx}
\usepackage{subfig}

\begin{document}
\newcommand{\room}{\;\;\,}
\newcommand{\Tr}{\mbox{Tr$\;$}}
\newcommand{\del}{\partial}
\newcommand{\mc}{\mathcal}
\newcommand{\p}[2]{\frac{\partial #1}{\partial #2}}
\newcommand{\pp}[3]{\frac{\partial^2 #1}{\partial #2\partial #3}}
\renewcommand{\[}{\left[}
\renewcommand{\]}{\right]}
\renewcommand{\(}{\left(}
\renewcommand{\)}{\right)}
\newcommand{\trix}[1]{\(\begin{array}{#1}}
\newcommand{\notrix}{\end{array}\)}
\setlength{\unitlength}{1.5cm}

\definecolor{babyblue}{rgb}{0.54, 0.81, 0.94}
\definecolor{corn}{rgb}{0.98, 0.93, 0.36}

\raggedbottom
\begin{titlepage}
\title{ {Moduli Axions, Stabilizing Moduli and the {\\[.01cm]}Large Field Swampland Conjecture{\\[.01cm]} in Heterotic M-Theory}\\[.01cm]}
                       
\author{{   
   C\'edric Deffayet,$^{a}$  Burt A. Ovrut$^{b}$ 
   and Paul J. Steinhardt$^{c}$}\\[0.5cm]  
{${}^a$\it Laboratoire de Physique de l'Ecole Normale Sup\'erieure,}\\[0.001cm]  
{\it ENS, Universit\'e PSL, CNRS} \\[0.001cm] {\it Sorbonne Universit\'e, Universit\'e Paris Cit\'e,} \\[0.001cm] {F-75005 Paris, France} \\[.01cm]
{${}^b$\it Department of Physics, University of Pennsylvania }  \\[0.001cm]  
{\it Philadelphia, PA 19104, USA} \\[0.01cm]
{${}^c$\it Department of Physics, Princeton University} \\[.001cm]
{\it Princeton, New Jersey 08544, USA}\\[0.01cm]
{${}^d$\it Jefferson Physical Laboratory, Harvard University}\\[0.001cm]
{\it  Cambridge MA 02138 USA}\\[0.01cm]     
}
\date{}
\maketitle
\begin{abstract}
\noindent  We compute the F- and D-term potential energy for the dilaton, complex structure and K\"ahler moduli of realistic vacua of heterotic M-theory compactified on Calabi-Yau threefolds where, for simplicity, we choose $h^{1,1} = h^{2,1} = 1$. However, the formalism is immediately applicable to the ``universal'' moduli of Calabi-Yau threefolds with $h^{1,1}=h^{1,2} > 1$ as well. The F-term potential is computed using the non-perturbative complex structure, gaugino condensate and “worldsheet instanton” superpotentials in theories in which the hidden sector contains an anomalous $U(1)$ structure group. The Green-Schwarz anomaly cancellation induces inhomogeneous ``axion'' transformations for the imaginary components of the dilaton and K\"ahler modulus  --  which then produce a D-term potential. $V_D$ is a function of the real components of the dilaton and K\"ahler modulus ($s$ and $t$) that is minimized and precisely vanishes along a unique line in the $s$-$t$ plane.   Excitations transverse to this line have a mass $m_{\rm anom}$ which is an explicit function of $t$. The F-term potential energy is then evaluated along the $V_{D}=0$ line. For values of $t$ small enough that $m_{\rm anom} \gtrsim M_{U}$ -- where $M_{U}$ is the compactification scale -- we plot $V_{F}$ for a realistic choice of coefficients as a function of the Pfaffian parameter $p$. We find values of $p$ for which $V_F$ has a global minimum at negative or zero vacuum density or a metastable minimum with positive vacuum density.  In all three cases,
the $s$, $t$ and associated ``axion'' moduli are completely stabilized. Finally, we show that, for any of these vacua, the large $t$ behavior of the potential energy satisfies the “large scalar field” Swampland conjecture.
\noindent 
\let\thefootnote\relax\footnotetext{\noindent cedric.deffayet@phys.ens.fr, ovrut@elcapitan.hep.upenn.edu, steinh@princeton.edu}
\end{abstract}

\thispagestyle{empty}
\end{titlepage}

\section{Introduction}

In this paper, we calculate and discuss the potential energy 
for the complex structure moduli, the K\"ahler moduli and the dilaton  in heterotic M-theory vacua \cite{Horava:1996ma, Horava:1995qa,Lukas:1998yy, Lukas:1998tt} whose hidden sector contains a line bundle with an anomalous $U(1)$ structure group  appropriately  embedded into $E_{8}$ \cite{Ashmore:2020ocb}. The Green-Schwarz cancellation \cite{Green:1984sg} of this $U(1)$ anomaly induces an inhomogeneous transformation of the imaginary components of the dilaton and each of the complex K\"ahler moduli. Thus, these imaginary components act as moduli ``axions''. The formalism introduced here is valid for any Calabi-Yau threefold with $h^{1,1}=h^{2,1} \geq 1$. However, for simplicity, we present our specific results for Calabi-Yau threefolds $X$ with $h^{1,1}=h^{2,1}=1$. These results are immediately applicable to the ``universal'' K\"ahler and complex structure moduli of Calabi-Yau compactifications with $h^{1,1}=h^{2,1}>1$ as well. We note that moduli stabilization has been discussed in a number of papers whose contexts differ from the present work -- see \cite{Gukov:2003cy, Balasubramanian:2005zx, Anderson:2011cza, Cicoli:2013rwa} for example.

The K\"ahler potential, $\mathcal{K}$, and superpotential, $W_{flux}$, for the complex structure moduli of a Calabi-Yau threefold with any $h^{2,1}\geq 1$ are well-known for ``large'' values of these moduli \cite{Candelas:1990pi, Gray:2007qy}. However, these quantities are not known analytically for ``small'' values.  Hence, for specificity, we will conduct our analysis using the large modulus formalism. We do not expect this to change the basic results of the potential energy calculation. In Section 2,  assuming $h^{2,1}=1$, we compute the $F$-term potential energy $V_{flux}$ for the complex structure modulus alone and show that there exists an infinite number of local minima that do {\it not} spontaneously break $N=1$ supersymmetry. Our complete extended calculation including both the dilaton and K\"ahler modulus will be carried out under the assumption that the complex structure modulus has been {\it fixed} to any one of these supersymmetric minima.

Having done this, in Section 3 we extend the calculation of $V_{F}$ to include the dilaton. This is accomplished under the assumption that the commutant of the anomalous $U(1)$ structure group contains a non-Abelian group that becomes strongly coupled at a high scale, thus leading to gaugino condensation \cite{Dine:1985rz, Nilles:1990zd, Horava:1996vs, Lukas:1997rb}. This produces a non-perturbative superpotential $W_{G}$ which, to lowest order, depends only on the complex dilaton field $S$. Using this, and the associated dilaton Kahler potential $K_{S}$, we extend our calculation of the $V_{F}$ potential energy using $W=W_{flux}+W_{G}$. 

The cohomology $H^{1,1}$ is associated with the homology group $H_{2}$.  For $h^{1,1}=1$, $H_{2}$ contains a single homology class. We denote this by $[C]$, where $C$ is an isolated, genus-zero holomorphic curve. It is well-known that the string can wrap itself around such a curve, producing a non-perturbative ``instanton'' superpotential \cite{Dine:1986zy, Dine:1987bq, Witten:1996bn} given by an exponential of the K\"ahler modulus $T$ multiplied by the ``Pfaffian'' associated with the Dirac operator. The Pfaffian is a holomorphic function of a subset of the vector bundle moduli evaluated at the curve $C$ \cite{Buchbinder:2002ic, Curio:2009wn}. Generically, $C$ is not unique -- with the number of such isolated, genus-zero curves in $[C]$ given by the Gromov-Witten invariant. As shown by Beasley and Witten \cite{Beasley:2003fx}, under a range of circumstances, the instanton superpotentials generated by all such curves can cancel exactly. However, for a wide set of vacua \cite{Buchbinder:2017azb, Buchbinder:2018hns, Buchbinder:2019hyb, Buchbinder:2019eal}, such as those whose Calabi-Yau threefold has a finite isotropy group \cite{Buchbinder:2016rmw}, this cancellation does not occur. Such theories then have an additional non-perturbative superpotential which is the sum of the instanton potentials over all isolated curves. We denote this superpotential, which depends on $T$, as $W_{T}$. Using this, and the associated K\"ahler potential $K_{T}$, in Section 4 we extend our calculation of the potential energy $V_{F}$ further by using $W=W_{flux}+W_{G}+W_{T}$.

Having computed the $F$-term potential, we recognize that the inhomogeneous transformation of the $S$ and $T$ moduli ``axions'' under the anomalous $U(1)$ structure group \cite{Dumitru:2021jlh, Dumitru:2022apw} leads to a second contribution to the potential energy - a $D$-term potential $V_{D}$ -- that must be added to $V_F$ to obtain the total potential energy $V$. The exact form of $V_D$ was derived in previous work \cite{Dumitru:2021jlh, Freedman:2012zz}. In Section 5 we present this potential and show that it is minimized, and $N=1$ supersymmetry is left unbroken, in a specific direction in field space with $s={\rm const.} \times t$ -- where $s={\rm Re} \, S, t={\rm Re}\, T$ and the constant is a fixed function of various parameters of the chosen vacuum. Along this direction, described as setting the Fayet-Iliopoulos term $FI=0$, the potential $V_D$ is minimized and exactly vanishes.
 For any  given pair of  $\langle s \rangle$ and $\langle t \rangle$ satisfying $FI=0$, we can expand both $S$ and $T$ around the associated local minimum. As discussed in detail in \cite{Dumitru:2021jlh, Dumitru:2022apw}, the complex fluctuations around this minimum can be unitarily rotated to two new complex fields that  have canonical kinetic energy and are mass eigenstates. We show that one of these new complex fields has a mass  $m_{\rm anom}$, that is, the mass of the anomalous $U(1)$ gauge field \cite{Ashmore:2020ocb}. For sufficiently small values of $\langle t \rangle$, this mass satisfies $m_{\rm anom} \gtrsim M_{U}$,  where $M_{U}$ is the compactification scale. Hence, both the gauge field and this diagonalized complex scalar can be  integrated out of the low energy theory. However, the second diagonalized complex scalar, with real and imaginary components $\eta$ and $\phi$ respectively, has canonical kinetic energy and, with respect to $V_{D}$ only, has vanishing mass. When $V_F$ is included, each of $\eta$ and $\phi$  get a non-vanishing mass {\it substantially smaller} than $M_{U}$. Hence, they remain in the low energy effective theory. In Section 5 we present the expressions for these lower mass scalars.

In Section 6, we now combine the results from both $V_F$ and $V_{D}$ and search for minima.  We first impose the relation $s =const. \times t$ along which $V_{D}$  is minimized and equal to zero;  then we  search for local extrema of $V_{F}$  along that specific direction. For sufficiently small values of $t$, where $m_{\rm anom}\gtrsim M_{U}$, we insert the results for the single light complex scalar field derived in the previous Section. Even with this reduced number of scalar fields, $V_{F}$ is a complicated function which we plot numerically using Mathematica. 
We find that the $V_{F}$  potential can have a unique minimum along  $s =const. \times t$  for a wide range of input parameters.  Specifically, we find that the potential energy can have a global minimum at $\langle t \rangle$ and, hence, $\langle s \rangle$ with a vanishing or negative cosmological constant or a metastable minimum with positive vacuum density -- depending on the explicit values of the input parameters. In this paper, for specificity, we choose an explicit set of physically motivated parameters and present the potential energy function for  various choices of the instanton Pfaffian parameter $p$. Within this context, we present a plot of $V_{F}$ for five different choices of $p$ which exhibit these characteristics. In addition to stabilizing the real components $s$, $t$ respectively, we show that the imaginary, axionic component of the light scalar is  simultaneously stabilized.

Finally, in Section 7, using the complete expression for $V_{F}$ valid at large values of $t$ where $m_{\rm anom} \ll M_{U}$, we show that $V_{F}$ easily satisfies the conjectured Swampland lower bound on the potential gradient \cite{Ooguri:2006in, Ooguri:2018wrx, Lust:2019zwm, Palti:2019pca, Bedroya:2019snp,Rudelius:2021azq}.    This is the case  for a wide range of values of Pfaffian parameter $p$, including cases where $V_F$ has a metastable minimum with positive vacuum energy.

In conclusion, we have shown that in heterotic M-theory vacua compactified on Calabi-Yau threefolds with $h^{1,1}=h^{2,1}=1$ and, more generically, for the universal K\"ahler and complex structure moduli of compactifications on any Calabi-Yau threefold with $h^{1,1}=h^{2,1}$,  the dilaton and geometric moduli can be completely stabilized. These vacua can have positive, zero or negative vacuum density depending on the precise parameters chosen for the theory. It is clear that these hidden sector vacua, in combination with the vacuum of the observable sector, will lead to a range of values for the cosmological constant. For hidden sector vacua that have positive, but small, vacuum density, one can expect a cosmological constant that can be adjusted to give the experimentally observed value. This will be discussed in detail in \cite{Paper2}. When redefined so as to have canonical kinetic energy, we find that the fluctuations around these vacua have positive definite masses with values sufficiently below the unification scale so as to allow these moduli fields to be in the low energy theory. Finally, we show that at large values of $t$, the Swampland ``large moduli field'' conjecture is satisfied -- even when the $V$ has a metastable minimum at small $t$ with positive vacuum density.

\section{Complex Structure Moduli}

We begin our analysis by considering the K\"ahler potential, superpotential and potential energy function for the complex structure moduli of a Calabi-Yau threefold $X$ within the context of heterotic $M$-theory. As discussed in the Introduction,  in this paper we will restrict our analysis to Calabi-Yau threefolds with cohomology $h^{2,1}=1$; that is, to a single complex structure modulus. However, it is instructive to begin our discussion by presenting the K\"ahler potential and superpotential for an arbitrary number of complex structure moduli, which we then restrict to the $h^{2,1}=1$ case.
\subsection{K\"ahler Potential}
The K\"ahler potential for an arbitrary number of complex structure moduli $z^{a},~a=1,\dots,h^{2,1} \geq 1$ was presented in \cite{Candelas:1990pi} and  given by
\begin{equation}
\kappa^{2}_{4}\mathcal{K}(z)=-ln\left[ 2i ({\cal{G}}-{\cal{\bar{G}}}) -i(z^{a}-{\bar{z}}^{a}) (\frac{\partial{\cal{G}}}{\partial{z^{a}}}+ \frac{\partial{\cal{\bar{G}}}}{\partial{\bar{z}^{a}}})  \right]
\label{1}
\end{equation}
\noindent where
 \begin{equation}
{\cal{G}}=-\frac{1}{6} {\tilde{d}}_{abc} z^{a}z^{b}z^{c} \ ,
 \label{2}
 \end{equation}
 $\tilde{d}_{abc}$ are the intersection numbers for the threefold $X$, 
 \begin{equation}
  \kappa^{2}_{4}=\frac{8\pi}{M_{P}^{2}}  
  \label{2a} 
  \end{equation}
  and
  \begin{equation}
  M_{P}=1.22 \times 10^{19} ~{\rm GeV} 
  \label{2b}
  \end{equation}
  is the unreduced Planck mass.

 Limiting the Calabi-Yau threefolds to those with $h^{2,1}=1$, that is, with a single complex modulus $z$ only, one can then write
 \begin{equation}
 \tilde{d}_{111}=6 {\tilde{d}}
 \label{3}
\end{equation}
where $\tilde{d}$ is a positive integer. The prepotential ${\mathcal{G}}$ then becomes
\begin{equation}
\mathcal{G}=-{\tilde{d}}z^{3} \ .
\label{4}
\end{equation}
It follows that the expression for the K\"ahler potential in \eqref{1} simplifies to
\begin{equation}
\kappa^{2}_{4}\mathcal{K}(z)=-ln\left[ i\tilde{d}(z-\bar{z})^{3} \right] \ .
\label{5}
\end{equation}

\subsection{Superpotential}

The flux generated superpotential in heterotic $M$-theory is given, at the classical level, by the Gukov-Vafa-Witten expression
\begin{equation}
W_{flux}=\frac{\sqrt{2}}{\kappa_{4}^{2}}\frac{1}{\pi \rho v^{1/2}} {\int_{X \times S^{1}/ \mathbb{Z}_{2}}} \Omega \wedge G \ ,   
\label{5a} 
\end{equation}
where $\Omega$ is the holomorphic three-form on the Calabi-Yau threefold $X$ and $G$ is the four-form  $G$-flux. Generically, $W_{flux}$ can depend on both the dilaton $S$  and the complex structure moduli $z^{a},~a=1,\dots,h^{2,1} \geq 1$. However, $W_{flux}$ was explicitly calculated within the context of heterotic $M$-theory in \cite{Gray:2007qy} and found to be independent of $S$ and given by 
\begin{equation}
W_{flux}={\cal{C}} ( \frac{1}{6}{\tilde{d}}_{abc}z^{a}z^{b}z^{c}n^{0}-\frac{1}{2}{\tilde{d}}_{abc}z^{a}z^{b}n^{c}-z^{a}n_{a}-n_{0} )
\label{6}
\end{equation}
where $n^0, n^a, n_0, n_a$ are units of flux and arbitrary independent integers in $\mathbb{Z}$,
\begin{equation}
{\cal{C}}=\frac{ \sqrt{2} v^{1/6} \epsilon_{0}}{\kappa^{2}_{4} (\pi \rho)^{2} }
\label{7}
\end{equation}
and
\begin{equation}
\epsilon_{0}= (2\sqrt{2}\pi)^{2/3} \frac{(\pi\rho)^{4/3}}{v^{1/3}M_{P}^{2/3}} \ .
\label{8}
\end{equation}
The parameters $v$ and $\pi\rho$ set the scale for the volume of the Calabi-Yau threefold $X$ and the length of the 5-th dimension respectively. We will choose these parameters as follows. First, let
\begin{equation}
v^{1/6}=\frac{1}{M_{U}} \ ,
\label{9}
\end{equation}
where $M_{U}$ is the unification scale of the effective low energy theory in the observable sector. As discussed in \cite{Deen:2016vyh, Ashmore:2020wwv}, a canonical value for $M_{U}$, which we will use in this paper, is given by
\begin{equation}
M_{U}=3.15 \times 10^{16}~ {\rm GeV} \ .
\label{10}
\end{equation}
Second, we set 
\begin{equation}
\pi\rho=5F v^{1/6}
\label{11}
\end{equation}
where $F$ is, at the moment, an arbitrary real number. To conform to standard values for $\pi\rho$ in the literature, see for example \cite{Banks:1996ss, Lukas:1997fg}, we will generically restrict
\begin{equation}
0.6 \lesssim F \lesssim 2 \ .
\label{12}
\end{equation}
However, as discussed in detail at the beginning of Section 6.4, both larger and smaller values for $F$ are acceptable for the specific vacua analyzed in this paper. Note that the ``physical'' volume of the Calabi-Yau threefold and the ``physical'' length of the 5-th dimension are given by
\begin{equation}
\mathbb{V}=vV ~~, ~~\mathbb{L}=\pi\rho\hat{R}
\label{13}
\end{equation}
where V and $\hat{R}$ are the moduli for the Calabi-Yau volume and fifth dimensional length respectively.
Using \eqref{9} and \eqref{11}, as well as the values of $M_{P}$ and $M_{U}$ given in \eqref{2b} and \eqref{10}, one can calculate the dimensionless coefficient $\epsilon_{0}$ in 
\eqref{8}. We find that
\begin{equation}
\epsilon_{0}=0.690 ~F^{4/3} \ .
\label{13a}
\end{equation}
Again using \eqref{2b},\eqref{9},\eqref{10},\eqref{11} as well as \eqref{13a}, we find that coefficient $\mathcal{C}$ in \eqref{7} is given by
\begin{equation}
\mathcal{C}= \frac{232}{F^{2/3}} ~M_{U}^{3} \ .
\label{13b}
\end{equation}
We have written $\mathcal{C}$  as proportional to $M_{U}^{3}$ to emphasize that $\mathcal{C}$ and, hence, $W_{flux}$ has mass dimension three set by the compactifiication scale. Finally, let us restrict the above superpotential to Calabi-Yau threefolds with $h^{2,1}=1$. Expression \eqref{6} then reduces to
\begin{equation}
W_{flux}={\cal{C}} ({\tilde{d}} z^{3}n^0 -3{\tilde{d}}z^2n^1-zn_1-n_0)
\label{14}
\end{equation}
where all coefficients remain unchanged and $n^0,n^1,n_0,n_1$ are independent, arbitrary elements of $\mathbb{Z}$. 

Before continuing on to the potential energy function for the complex structure modulus, we first compute the covariant derivative of $W_{flux}$ in \eqref{14} with respect to $z$ defined by
\begin{equation}
D_{z} W_{flux}=\frac{\partial W_{flux}}{\partial z}+\kappa^{2}_{4} \frac{\partial{\cal{K}}}{\partial z}W_{flux}.
\label{15}
\end{equation}
Using \eqref{5} and \eqref{14}, we find that
\begin{equation}
D_{z} W_{flux}=\frac{\cal{C}}{z-\bar{z}} \big( 3 \tilde{d} z^2 n^1 +2zn_{1} -3\tilde{d}z^2\bar{z}n^0 +6\tilde{d}z\bar{z}n^1 +n_1\bar{z} +3n_0  \big) .
\label{16}
\end{equation}
In order for a complex structure moduli vacuum to preserve $N=1$ supersymmetry, it must satisfy
\begin{equation}
D_{z} W_{flux}=0 
\label{17}
\end{equation}
and, hence, from \eqref{16} that
\begin{equation}
3 \tilde{d} z^2 n^1 +2zn_{1} -3\tilde{d}z^2\bar{z}n^0 +6\tilde{d}z\bar{z}n^1 +n_1\bar{z} +3n_0 =0 \ .
\label{18}
\end{equation}
Defining
\begin{equation}
z=r+ic \ ,
\label{19}
\end{equation}\
we find that this will be the case if and only if
%
%
\begin{equation}
r=-\frac{3 n_0n^0+n_1n^1}{2(3\tilde{d}(n^1)^2+n^0 n_1)} 
\label{20}
\end{equation}
and
\begin{equation}
c=\pm \big(-r^2 +\frac{2n^1r}{n^0}+\frac{n_1}{3 \tilde{d}n^0} \big)^{1/2} \ .
\label{21}
\end{equation}
There are, of course, an infinite number of solution to r and c of this form, indexed by the choices of integers $n^0,n^1,n_0,n_1$. As a simple example, let us consider solutions where 
\begin{equation}
r=0~~\Rightarrow~~ 3 n_0n^0=-n_1n^1~~{\rm and}~~c=\pm\big(\frac{n_1}{3\tilde{d}n^0}\big)^{1/2} \ .
\label{21a}
\end{equation}
Furthermore, by taking 
\begin{equation}
n_1=n^0=-n^1~~\Rightarrow~~c=\pm\frac{1}{\sqrt{3 \tilde{d}}}~~{\rm and}~~3 n_0=n_1=n^0=-n^1 \ .
\label{24b}
\end{equation}
Therefore, any choice of this one parameter set of integers gives a unique complex structure modulus, with $r=0$ and a simple value of $c$, that satisfies $D_{z} W_{flux}=0$.

As we proceed, it will become essential to know the value of $W_{flux}$ at a point where $D_{z} W_{flux}=0$. 
It follows from \eqref{14},\eqref{15},\eqref{17} and \eqref{5} that at a supersymmetric point
%
\begin{equation}
W_{flux}= ~\mathcal{C}\tilde{d}c
\big(-4(n^0r-n^1)c+i[2n^0(r^2-c^2)-4n^1r-\frac{2n_1}{3\tilde{d}} ]\big) .
\label{22}
\end{equation}
Using \eqref{21}, this simplifies to 
\begin{equation}
W_{flux}=-4{\mathcal{C}} \tilde{d} c^{2}\big( (n^0r-n^1)+in^0c \big) \ .
\label{23}
\end{equation}
We find it convenient to re-express this as
\begin{equation}
W_{flux}=-4{\mathcal{C}} \tilde{d}\big(A+iB\big) \ ,
\label{23a}
\end{equation}
where 
\begin{equation}
A=c^{2}(n^0r-n^1),~~B=c^3n^0,~~\mathcal{C}= \frac{232}{F^{2/3}} ~M_{U}^{3} 
\label{23b}
\end{equation}
and $r$ and $c$ are given in \eqref{20} and \eqref{21} respectively.
As a concrete example, consider the choice of integer parameters given in \eqref{24b}. Then 
%
\begin{equation}
r=0~,~c=\pm\frac{1}{\sqrt{3 \tilde{d}}} ~~~{\Rightarrow}~~~~A=\frac{n^0}{3\tilde{d}}~~,~B= \pm \frac{n^0}{3\tilde{d}\sqrt{3\tilde{d}}}~~~{\rm for}~n^0\in \mathbb{Z}  \ .
\label{23c}
\end{equation}

\subsection{Complex Structure F-Term Potential Energy}

In this subsection, we want to discuss the part of the F-term potential energy function arising from the flux superpotential $W_{flux}$ only. To do this, however, it is essential to include
contributions that arise from terms associated with the K\"ahler potentials of the dilaton $S$ and the $h^{1,1}=1$ K\"ahler modulus $T$, as well as the complex structure $z$. That is, the relevant terms in the potential energy function are
\begin{equation}
\begin{aligned}
V_{flux} & =e^{\kappa^{2}_{4}K}\big[K^{z\bar{z}} D_{z}W_{flux} D_{\bar{z}} \bar{W}_{flux} + K^{S\bar{S}} D_{S}W_{flux} D_{\bar{S}} \bar{W}_{flux} \\
             & +K^{T\bar{T}} D_{T}W_{flux} D_{\bar{T}} \bar{W}_{flux} - 3 \kappa^{2}_{4}|W_{flux}|^{2} \big] \ ,
\end{aligned}
\label{24}
\end{equation}
where
\begin{equation}
\begin{aligned}
\kappa^{2}_{4}K &=\kappa^{2}_{4}(K_{S}+K_{T}+\mathcal{K}) \\ 
\label{25}
\kappa^{2}_{4}K_{S}=-ln(S+\bar{S})~,~\kappa^{2}_{4}K_{T}&=-3ln(T+\bar{T})~, ~\kappa^{2}_{4}\mathcal{K}=-ln(i\tilde{d}(z-\bar{z})^{3})
\end{aligned}
\end{equation}
and 
%
%
\begin{equation}
D_{i}W_{flux}=\partial _{i}W_{flux}+\kappa^{2}_{4}(\partial_i K)W_{flux}~~{\rm for}~~i=S,T,z \ .
\label{26a}
\end{equation}
We emphasize that here, and {\it throughout this paper,} we always use the {\it tree level} expressions--given in \eqref{25}--for the K\"ahler potentials. This is motivated by the fact that, in the absence of five-branes--which we are assuming--the leading order string perturbative corrections vanish. See, for example, \cite{Dumitru:2022apw}. Evaluating $K^{S \bar{S}}$ and $K^{T\bar{T}}$ and using the fact that $W_{flux}$ is independent of both $S$ and $T$, expression \eqref{24} becomes
\begin{equation}
\begin{aligned}
V_{flux} & =e^{\kappa^{2}_{4}K}\big[K^{z\bar{z}} D_{z}W_{flux} D_{\bar{z}} \bar{W}_{flux} + \kappa^{2}_{4} |W_{flux}|^{2} \\
               & +3\kappa^{2}_{4} |W_{flux}|^{2} -3\kappa^{2}_{4} |W_{flux}|^{2}  \big ] \ ,
 \end{aligned}             
\label{27}
\end{equation}
which , noting the cancellation of the last two $|W_{flux}|^{2}$ terms, simplifies to
\begin{equation}
V_{flux} =e^{\kappa^{2}_{4}K}\big[K^{z\bar{z}} D_{z}W_{flux} D_{\bar{z}} \bar{W}_{flux} + \kappa^{2}_{4} |W_{flux}|^{2}  \big] \ .
\label{28}
\end{equation}
Finally, using
\begin{equation}
e^{\kappa^{2}_{4}K} = \frac{-i}{(S+\bar{S})(T+\bar{T})^{3}\tilde{d}(z-\bar{z})^{3}}~,~K^{z\bar{z}}= -\frac{\kappa^{2}_{4}(z-\bar{z})^{2}}{3} \ ,
\label{29}
\end{equation}
it follows that 
\begin{equation}
V_{flux} = \frac{1}{(S+\bar{S})(T+\bar{T})^{3}} 
\big[  \frac{-i}{\tilde{d}(z-\bar{z})^{3}}
\big( -\frac{\kappa^{2}_{4}(z-\bar{z})^{2}}{3}D_{z}W_{flux} D_{\bar{z}} \bar{W}_{flux} + \kappa^{2}_{4} |W_{flux}|^{2} \big) \big] 
\label{30}
\end{equation}
Note that this is the potential energy function for arbitrary values of complex structure modulus z. Hence, $D_{\bar{z}} \bar{W}_{flux}$ does not generically vanish.

Although complex structure moduli satisfying constraints \eqref{20} and \eqref{21} have vanishing covariant derivative $D_{z}W_{flux}$, it is essential to determine whether or not they are local minima of $V_{flux}$. We begin by considering the first derivative of the potential function \eqref{30}.
 We find that
\begin{equation}
\frac{\partial V_{flux}}{\partial{z}}=\frac{-i\kappa^{2}_{4}}{(S+\bar{S})(T+\bar{T})^{3}\tilde{d}(z-\bar{z})^{3}} \big[ \frac{\partial W_{flux}}{\partial z}-\frac{3}{(z-\bar{z})} W_{flux}\big] \bar{W}_{flux}
\label{31}
\end{equation}
Using \eqref{5} and \eqref{15}, it follows that
\begin{equation}
 \frac{\partial W_{flux}}{\partial z}-\frac{3}{(z-\bar{z})} W_{flux}=D_{z}W_{flux}
 \label{32}
 \end{equation}
 and, hence, 
 \begin{equation}
 \frac{\partial V_{flux}}{\partial{z}}=\frac{-i\kappa^{2}_{4}}{(S+\bar{S})(T+\bar{T})^{3}\tilde{d}(z-\bar{z})^{3}} \big[D_{z}W_{flux}\big] \bar{W}_{flux}\ .
 \label{32b}
 \end{equation}
 Therefore, all complex structure moduli with vanishing covariant derivative $D_{z}W_{flux}=0$ satisfy
\begin{equation}
\frac{\partial V_{flux}}{\partial{z}}=0 \ .
\label{33}
\end{equation}
Since $V_{flux}$ is real, this implies that
\begin{equation}
\frac{\partial V_{flux}}{\partial{r}}=\frac{\partial V_{flux}}{\partial{c}}=0 \ .
\label{34}
\end{equation}
Are these extrema with $D_{z}W_{flux}=0$ also local minima of $V_{flux}$? The answer is yes, {\it provided  $c  >0$.}  This can be proven by a tedious but straightforward evaluation of the second derivatives of $V_{flux}$ at the extrema which shows that    
 \begin{equation}
\frac{\partial^{2}V_{flux}}{\partial^{2}r}>0, \frac{\partial^{2}V_{flux}}{\partial^{2}c}>0, \, {\rm and} \,\frac{\partial^{2}V_{flux}}{\partial r\partial c} =0 
\, ~{\rm for}~  \langle  c \rangle >0.
\label{35}
\end{equation}
This suffices to prove that the determinant of the Hessian and the second derivative  $V_{flux}$ with respect to $r$ are both positive at the extrema and, therefore, the extrema must be minima.

Henceforth, we will chose the vacuum expectation value $\langle z \rangle$ of the complex structure modulus to be at one of the local minima of $V_{flux}$. It follows from \eqref{23a},\eqref{23b} and \eqref{30} that, at any such  minimum, 
\begin{equation}
V_{flux}=\frac{M_{U}^{4}}{st^{3} \langle  c \rangle^3} \big(\frac{1.14}{F^{4/3}}\big)\tilde{d}(A^{2}+B^{2}) \ .
\label{35a}
\end{equation}

 \subsection{Non-Renormalization of $W_{flux}$}

The Gukov-Vafa-Witten formula in \eqref{5a} gives the tree level expression for $W_{flux}$. Can $W_{flux}$ receive string and $\alpha^{\prime}$ perturbative corrections? Within the context of heterotic $M$-theory the answer is no. As shown in \cite{Burgess:2005jx}, low energy Peccei-Quinn symmetry forbids $W_{flux}$ from containing the dilaton $S$ both at the tree level--thus explaining why $S$ does not appear in \eqref{6}--and at any perturbative level--thus ensuring that $W_{flux}$ receives no corrections in string perturbation theory. Similarly, axion-like symmetries of the K\"ahler moduli preclude $W_{flux}$ from acquiring a $T$-dependence. Hence, $W_{flux}$ can receive no $\alpha^{\prime}$ corrections.  That is, the expression for $W_{flux}$ in \eqref{6} can receive no string and $\alpha^{\prime}$ perturbative corrections.

Be that as it may, these arguments no longer apply to non-perturbative corrections to $W_{flux}$, to which we now turn.

\section{Gaugino Condensation and $N=1$ Supersymmetry Breaking}

To spontaneously break $N=1$ supersymmetry, following, for example, the approach in the heterotic M-theory $B-L$ MSSM vacuum \cite{Ashmore:2020ocb,Braun:2005nv}, we consider theories containing a line bundle with an anomalous $U(1)$ structure group in the hidden sector. Furthermore, we assume that the commutant subgroup of $U(1)$  contains a non-Abelian group which becomes strongly coupled  at a mass scale $\Lambda$ near $M_{U}$. 

\subsection{Gaugino Condensate Superpotential}
As is well-known \cite{Dine:1985rz, Nilles:1990zd, Horava:1996vs, Lukas:1997rb}, the condensation of the associated gauginos at scale $\Lambda$ produces a non-perturbative superpotential which, to lowest order, is given by
\begin{equation}
W_{G}=M_{U}^{3} e^{-bS}   \ ,
\label{36}
\end{equation}
where
\begin{equation}
b=\frac{6\pi}{b_{L}{\hat{\alpha}}_{GUT}}~~,~~{\hat{\alpha}}_{GUT}=\frac{\big(8\pi(\pi\rho)\kappa^{2}_{4}\big)^{2/3}}{v^{1/3}} 
\label{37}
\end{equation}
and $b_{L}$ is the renomalization group coefficient associated with the specific non-Abelian commutant. Using \eqref{2a},\eqref{9},\eqref{10} and \eqref{11}, we find that
\begin{equation}
{\hat{\alpha}}_{GUT}=0.0762 ~F^{2/3}
\label{38}
\end{equation}
where $F$ can run over the range given in \eqref{12}. The value of $b_{L}$ depends on the specific commutant non-Abelian group in the hidden sector. For example, in \cite{Ashmore:2020ocb,Ashmore:2020wwv}, the non-Abelian group is $E_{7}$ with the renormalization group coefficient
\begin{equation}
b_{L}=6 \ .
\label{39}
\end{equation}
The non-perturbative superpotential $W_{G}$ leads to spontaneous breaking of $N=1$ supersymmetry in the $S$ and $T$ moduli, which is then gravitationally mediated to the observable matter sector. The scale of SUSY breaking in the low-energy observable sector is of order
\begin{equation}
m_{\rm susy} \sim \kappa_{4}^{2}\Lambda^{3}=8\pi\frac{\Lambda^{3}}{M_{P}^{2}} \ .
\label{39a}
\end{equation}

Let us now extend the superpotential  to include {\it both} the complex structure  and gaugino condensate superpotentials. That is, take
\begin{equation}
W=W_{flux}+W_{G} \ .
\label{40}
\end{equation}

\subsection{F-Term Potential Including Gaugino Condensation}
Generically, the F-term potential energy function is given by
\begin{equation}
V_{F}=e^{\kappa^{2}_{4}K}\big[ K^{i\bar{j}}D_{i}W D_{\bar{j}}{\bar{W}}-3\kappa^{2}_{4}|W|^{2} \big]
\label{41}
\end{equation}
where 
\begin{equation}
D_{i}W=\partial_{i}W+\kappa_{4}^{2}(\partial_{i}K)W \ ,
\label{42}
\end{equation}
the K\"ahler potential is given in \eqref{25} and the indices $i,j$ run over $S,T$ and $z$. In this Section, we choose the superpotential to be that given in \eqref{40}.

In our evaluation of $V_{F}$ we will, henceforth, assume that the complex structure modulus $z$ is always fixed to be a local minimum $\langle z \rangle$ of $V_{flux}$ satisfying $D_{z}W_{flux}=0$.
For example, the evaluation of the $z$ covariant derivative in \eqref{41} is given by
\begin{equation}
\begin{aligned}
D_{z}W & =\partial_{z}(W_{flux}+W_{G})+\kappa_{4}^{2} \partial_{z}\mathcal{K}(W_{flux}+W_{G}) \\
& =D_{z}W_{flux} +(\frac{-3}{(z-\bar{z})}) W_{G} \\
& =\frac{-3}{\langle z-{\bar{z}} \rangle}W_{G} \ ,
\label{43}
\end{aligned}
\end{equation}
where we have used the fact that the gaugino condensate superpotential \eqref{36} is independent of complex structure.
Then, using $W_{flux}$ given in \eqref{23a}, \eqref{23b}, the gaugino condensate superpotential $W_{G}$ in \eqref{36}, the K\"ahler potentials in \eqref{25} and defining
\begin{equation}
S=s+i\sigma~~,~~T=t+i\chi
\label{44}
\end{equation}
we find, after a moderate calculation, that
\begin{equation}
\begin{aligned}
V_{F} =&\frac{M_{U}^{4}}{st^{3}\langle c \rangle^{3}} \big[( \frac{1.1376}{F^{4/3}})~ \tilde{d}(A^2+B^2) \\
+&1.32 \times 10^{-6}\tilde{d}^{-1}\big((1+2bs)^{2}+3 \big)e^{-2bs}  \\
- &(\frac{2.43 \times 10^{-3}}{F^{2/3}} )\big(1+2bs\big)e^{-bs} \big(A \ \cos(b\sigma)-B \sin(b\sigma)\big)  \big] \ .
\label{45}
\end{aligned}
\end{equation}
Once again, note that we are using the tree level expressions for the K\"ahler potentials given in \eqref{25}.
 
We conclude that adding the non-perturbative superpotential $W_{G}$ substantially alters the F-term effective potential $V_{F}$  presented in \eqref{35a} based on $W_{flux}$ alone and evaluated at 
$D_{z}W_{flux}=0$. It is important, therefore, to search for and include any other non-perturbative superpotentials in the calculation of the F-term potential energy.

\section{Worldsheet Instanton Superpotential}

It is well-known \cite{Dine:1986zy, Dine:1987bq, Witten:1996bn}  that a non-perturbative contribution to the superpotential can also be generated by the superstring wrapping around isolated, genus zero, holomorphic curves in the Calabi-Yau compactification threefold. The leading order contribution to this superpotential arises from the superstring wrapping {\it once} around each such curve. In this Section, we introduce this leading order, non-perturbative ``worldsheet instanton'' contribution.

\subsection{Single Isolated, Genus-Zero Curve}

Let $C$ be an holomorphic, isolated, genus zero curve in the Calabi-Yau threefold $X$ with $h^{1,1}=1$. Then, as discussed in \cite{Buchbinder:2003pi}, the general form of the instanton superpotential induced by a string wrapping $C$ is given by
\begin{equation}
W_{I}=\mathcal{P}e^{-\tau T} \ ,
\label{46}
\end{equation}
where 
\begin{equation}
\tau=\frac{1}{2}T_{M}(\pi\rho)v_{C} \ ,
\label{48}
\end{equation}
$v_{C}$ is the volume of the holomorphic curve $C$
and $T_{M}$ is the string membrane tension
\begin{equation}
T_{M}=\frac{1}{(v \rho\kappa^{2}_{4})^{1/3}} \ .
\label{49}
\end{equation}
Using \eqref{2a},\eqref{9},\eqref{10},\eqref{11} one finds 
\begin{equation}
T_{M}=\frac{15.5}{F^{1/3}}~M_{U}^{3}
\label{50}
\end{equation}
and, hence, that
\begin{equation}
\tau=38.8~F^{2/3} \frac{v_{C}}{v^{1/3}}
\label{51}
\end{equation}

The factor $\mathcal{P}$ in \eqref{46} is given by
\begin{equation}
\mathcal{P} \propto {\rm Pfaff}(\mathcal{D}_{-}) \ ,
\label{52}
\end{equation}
where ${\rm Pfaff}(\mathcal{D}_{-})$ is the Pfaffian of the chiral Dirac operator constructed using the hermitian Yang-Mills connection associated with the holomorphic vector bundle  on both the observable and hidden sectors evaluated at the curve $C$. For a generic bundle, one expects the Pfaffian to be proportional to a holomorphic, polynomial function of a subset of the vector bundle moduli, as shown in several contexts in \cite{Buchbinder:2002ic, Curio:2009wn, Buchbinder:2016rmw}. In the previous Section, we specified that the vector bundle in the hidden sector must contain a line bundle with an anomalous $U(1)$ structure group. If the hidden sector bundle is strictly a single line bundle $L$ (more exactly, its extension to $L \oplus L^{-1}$ whose structure group is embedded in $E_{8}$), we note that it will have no vector bundle moduli and, hence, vanishing Pfaffian. However, it is important to note that the ``string instanton'' arises from all bundle gauge connections in both the hidden and the observable sectors. That is, $\mathcal{P}$ is the Pfaffian associated with the gauge connections of the vector bundles of both the observable and hidden sectors, which, as a rule, will have multiple vector bundle moduli and, potentially, a non-vanishing Pfaffian. Finally, we note that the proportionality factor in \eqref{52} was defined in detail, for example in \cite{Buchbinder:2016rmw, Curio:2010hd}, and shown to be an explicit function of the complex structure moduli. In this paper, since the value of the complex structure modulus has been fixed to be at a supersymmetry preserving minimum of $V_{flux}$, this function becomes a constant that could, in principle, be explicitly computed. However, for simplicity, we will simply absorb it into the expression for the Pfaffian and write 
\begin{equation}
\mathcal{P}={\rm Pfaff}{\mathcal({\mathcal{D}}}_{-} ) \ .
\label{53}
\end{equation}

\subsection{Multiple Isolated, Genus-Zero Curves}

When the Calabi-Yau threefold has $h^{1,1}=1$, there is a single homology class in $H_{2}$ which,  given the above discussion, we denote by $[C]$. It is well known that, in general, this class can contain a finite number of holomorphic, isolated, genus zero curves in addition to $C$. The total number of such curves is specified by the Gromov-Witten invariant, which can be computed given a specific Calabi-Yau threefold -- see, for example \cite{Buchbinder:2016rmw, Braun:2007xh,Braun:2007tp}. Denote the  number of such curves by $n_{[C]}$ and write them as $C_{i},~ i=1,\dots n_{[C]}$.
All such curves in the same homology class have the same area, the same classical action and the same exponential factor as in \eqref{46}. However, their Pfaffians are, in general, different. Hence, the contribution to the superpotential from all curves $C_{i}$ in the homology class $[C]$ is given by
\begin{equation}
W_{I}= \big( \sum_{i=1}^{n_{[C]}} \mathcal{P}_{i}\big)e^{-\tau T} \ .
\label{54}
\end{equation}

An important theorem of Beasley and Witten \cite{Beasley:2003fx} proved that, under a range of circumstances, the sum over the Pfaffians in any given homology class must vanish. It then follows that the associated instanton superpotential in \eqref{54} would be zero. However, as shown in a number of papers -- see \cite{Buchbinder:2017azb,Buchbinder:2018hns, Buchbinder:2019hyb, Buchbinder:2019eal} for example -- there are a substantial number of phenomenologically acceptable vacua that violate one or more of the assumptions in the Beasley-Witten theorem. For example, in the heterotic M-theory $B-L$ MSSM vacuum in \cite{Buchbinder:2016rmw}, the 
existence of $\mathbb{Z}_{3} \times \mathbb{Z}_{3}$ discrete torsion violates the Beasley-Witten theorem and has non-vanishing instanton superpotential $W_{I}$. Henceforth, we will assume that the vacua we are considering violate the Beasley-Witten theorem; that is, that $ \sum_{i=1}^{n_{[C]}} \mathcal{P}_{i} \neq 0$. 
%
%

The exact functional form of $\sum_{i=1}^{n_{[C]}} \mathcal{P}_{i}$ has only been calculated in a few specific examples, see  \cite{Buchbinder:2002ic, Curio:2009wn, Buchbinder:2016rmw}, and is unknown for most physically acceptable heterotic vacua. This is also the case for the vector bundle moduli K\"ahler potential. Hence, the precise formalism for computing the supersymmetry preserving vector bundle moduli vacua is unknown. Be that as it may, 
we will now assert--similarly to the complex structure modulus--that {\it the values of the vector bundle moduli are independently fixed to be constants in a vacuum state that does not break $N=1$ supersymmetry}. It follows that the Pfaffian factor is simply a complex number, which we will express as 
\begin{equation}
\sum_{i=1}^{n_{[C]}} \mathcal{P}_{i} = p e^{i\theta_{p}} \ ,
\label{home1}
\end{equation}
where, since it  is unknown how to compute the Pfaffian for the theories discussed in this paper, the values of $p$ and $\theta_{p}$ are, a priori, unrestricted.
That is, in this paper, we simply parameterize the Pfaffian factor in terms of the real coefficients $p$ and $\theta_{p}$ and deduce what their values must be in order to stabilize the dilaton, real K\"ahler modulus and the $T$ axion respectively  within our context.

In summary, we will henceforth assume that $W_{I} \neq 0$, and simply denote \eqref{54} as
\begin{equation}
W_{I}= M_{U}^{3}pe^{i\theta_{p}}e^{-\tau T} \ ,
\label{55}
\end{equation}
where, using the fact that the Calabi-Yau threefold in this paper has a mass scale of order $M_{U}$, we have restored natural units.

\subsection{F-Term Potential Function With String Instantons}

In this subsection, we extend the F-term potential energy function given in \eqref{45} to include contributions from the instanton superpotential. That is, we evaluate the potential energy function in \eqref{41},\eqref{42} but now taking the superpotential to be
\begin{equation}
W=W_{flux}+W_{G}+W_{I} \ ,
\label{56}
\end{equation}
where $W_{flux}$, $W_{C}$ and $W_{I}$ are given in \eqref{23a}, \eqref{36} and \eqref{55} respectively. As above, the K\"ahler potential is presented in \eqref{25} and the indices $i,j$ run over $S$, $T$ and $z$. We emphasize once again that throughout this paper we always use the tree level expressions for the K\"ahler potentials. Importantly, as we did in the previous Section, we will assume  in our evaluation of $V_{F}$ that the complex structure $z$ is always fixed to be a local minimum $\langle z \rangle$ of $V_{flux}$ satisfying $D_{z}W_{flux}=0$. Again defining
\begin{equation}
S=s+i\sigma~~,~~T=t+i\chi
\label{57}
\end{equation}
we find, after a detailed calculation, that
\begin{equation}
\begin{aligned}
V_{F}=&\frac{M_{U}^{4}}{st^{3}\langle c \rangle^{3}} \big[( \frac{1.14}{F^{4/3}})~ \tilde{d}(A^2+B^2) \\
+&1.32 \times 10^{-6}\tilde{d}^{-1}\big((1+2bs)^{2}+3 \big)e^{-2bs}  \\
-&(\frac{2.43 \times 10^{-3}}{F^{2/3}} )\big(1+2bs\big)e^{-bs} sgn(A)\sqrt{A^2+B^2} \cos(b\sigma+ \arctan(\frac{B}{A})) \\
+&2.62 \times 10^{-6} p\tilde{d}^{-1} \big(1+2bs+3(\tau t +\frac{3}{2})\big)e^{-bs-\tau t}  \cos(b\sigma-\tau\chi+\theta_{p})\\
+&4.36 \times 10^{-7} p^{2}\tilde{d}^{-1}\big(3+(2\tau t+3)^{2}\big)e^{-2\tau t}  \\
-&(\frac{2.43  \times 10^{-3}}{F^{2/3}} ) p (1+2\tau t)e^{-\tau t} sgn(A)\sqrt{A^2+B^2} \cos(\tau \chi-\theta_{p}+ \arctan(\frac{B}{A}))  \big]  \ ,
\end{aligned}
\label{58}
\end{equation}
where $A$, $B$, $\langle c \rangle$ are defined in \eqref{23b}, F is defined in \eqref{11},\eqref{12}, b and $\tau$ are given in \eqref{37} and \eqref{51}, respectively, and $p$ and $\theta_{p}$ arise in \eqref{55}. In addition, we have used the trigonometric relation
\begin{equation}
A \cos(x)-Bsin(x)=sgn(A)\sqrt{A^2+B^2} \cos(x+ \arctan(\frac{B}{A})) \ .
\label{59}
\end{equation}

We conclude that adding the ``string instanton'' superpotential $W_{I}$ greatly alters the F-term potential energy function -- which, for $W_{flux}+W_{G}$ was presented in \eqref{45}. The potential $V_{F}$ in \eqref{58} is -- to lowest order -- the most general expression for the F-term potential energy for the dilaton and K\"ahler modulus when the complex structure $z$ is evaluated at a local minimum of $V_{flux}$ for which $D_{z}W_{flux}=0$.

\section{Anomalous $U(1)$ and $\sigma$,$\chi$ Axions}

In the previous Sections, we have considered only the superpotentials and the F-term potential energy associated with the dilaton, complex structure and K\"ahler moduli. This was motivated by the fact that none of these moduli fields transform homogeneously under any low-energy gauge group. However, as we discuss in this Section, the anomalous $U(1)$ gauge group does produce an {\it  inhomogeneous} transformation of the imaginary component of both the dilation and the K\"ahler modulus -- thus inducing a D-term potential energy $V_{D}$. In this Section, we define and discuss this D-term potential only. In Section 6, we will reintroduce $V_{F}$ in combination with $V_{D}$ and discuss the associated moduli vacua.

\subsection{Inhomogeneous $U(1)$ Transformations}
As discussed in Section 2, in this paper we will assume that the hidden sector vector bundle contains a line bundle $L$ with an anomalous $U(1)$ structure group. Within the context of compactification on a Calabi-Yau threefold with $h^{1,1}=1$, it was shown in \cite{Dumitru:2022apw} that the Green-Schwarz mechanism \cite{Green:1984sg} cancels this anomaly by producing an inhomogeneous transformation of the dilaton $S$ and the K\"ahler modulus $T$.  It was shown in \cite{Dumitru:2021jlh, Dumitru:2022apw} that for a $U(1)$ parameter $\theta$, these transformations are given by \\

\begin{equation}
\delta_{\theta}S = 2i\pi a \epsilon_{S}^{2} \epsilon_{R}^{2} \beta l \theta \equiv k_{S}\theta ~~,~~
\delta_{\theta}T=-2i a\epsilon_{S}\epsilon^{2}_{R}l \theta \equiv k_{T} \theta
\label{60}
\end{equation}
where
\begin{equation}
\epsilon_{S}=\frac{2\pi \rho^{4/3}}{v^{1/3}M_{P}^{2/3}}~~,~~ \epsilon_{R}=\frac{v^{1/6}}{\pi \rho}
\label{61}
\end{equation}
are strong coupling expansion parameters. The integer $l$ defines the line bundle as $\mathcal{O}_{X}(l) =L$,  $\beta$ is the gauge charge on the hidden sector and
\begin{equation}
a=\frac{1}{4}tr_{E_{8} }Q^{2}
\label{62}
\end{equation}
is determined given the embedding matrix $Q$ of $U(1)$ into $E_{8}$ in the hidden sector.

Using \eqref{2a}, \eqref{9}, \eqref{10} and \eqref{11} one can determine the expansion parameters in \eqref{61}. We find that
\begin{equation}
\epsilon_{S}=0.220~F^{4/3}~~,~~\epsilon_{R}= \frac{0.2}{F} \ .
\label{63}
\end{equation}
Note that the $\epsilon_{S}$ often occurs multiplied by $\pi$, so we introduce
\begin{equation}
\epsilon^{\prime}_{S}=\epsilon_{S}\pi=0.690~F^{4/3} \ .
\label{64}
\end{equation}
For simplicity, we will assume henceforth that the  line bundle structure group embeds into the $SU(2)$ subgroup of $E_{8}$ in the hidden sector. It follows that $a=1$. Finally, we will defer a discussion of the values of $l$ and $\beta$ until later in the paper.

We see from \eqref{60} that these $U(1)$ transformations are purely imaginary and, hence, represent inhomogeneous transformations of the $\sigma$ and $\chi$ imaginary components of $S$ and $T$ respectively. That is, the $\sigma$ and $\chi$ fields behave as  ``axions'' under the anomalous $U(1)$ transformation.

\subsection{Anomalous $U(1)$ Induced Potential Energy $V_{D}$}

In addition to the canonical D-term potential energy involving all hidden sector matter fields carrying  non-vanishing $U(1)$ charge, the inhomogeneous $U(1)$ transformations of $S$ and $T$ presented in \eqref{60} induce a potential energy for the dilaton and modulus fields even though they are neutral under linear $U(1)$ transformations. As discussed in detail in \cite
{Dumitru:2022apw}, assuming the vacuum expectation values for all hidden matter scalars are zero, then these matter scalars ``decouple'' from the $S$ and $T$ moduli D-term potential energy. We will assume that this is the case in this paper. Then, as shown in \cite{Dumitru:2022apw, Freedman:2012zz}, ignoring the matter scalars, the D-term potential energy is given by
\begin{equation}
V_{D}=\frac{1}{2s}\mathcal{\mathbb{P}}^{2} \ ,
\label{65}
\end{equation}
where
\begin{equation}
\mathbb{P}=ik_{S}\partial_{S}K+ik_{T}\partial_{T}K \ .
\label{66}
\end{equation}
It follows from \eqref{25} that
\begin{equation}
\partial_{S}K=-\kappa^{-2}_{4}\frac{1}{S+\bar{S}} =-\kappa^{-2}_{4}\frac{1}{2s}~~, ~~\partial_{T}K=-3\kappa^{-2}_{4}\frac{1}{T+\bar{T}} =-3\kappa^{-2}_{4}\frac{1}{2t}
\label{67}
\end{equation}
and from \eqref{60} that
\begin{equation}
k_{S}=i2\pi\epsilon_{S}^{2}\epsilon_{R}^{2}\beta l ~~~,~~~k_{T}=-i 2 \epsilon_{S}\epsilon_{R}^{2} l \ .
\label{68}
\end{equation}
Inserting these results into \eqref{66}, we find that
\begin{equation}
\mathbb{P}=-\frac{\epsilon_{S}\epsilon_{R}^{2}}{\kappa^{2}_{4}} \Big( -\frac{1}{s} \epsilon^{\prime}_{S}\beta l +\frac{3l}{t} \big) \ .
\label{69}
\end{equation}
Using \eqref{63}, \eqref{64} as well as \eqref{2a} and \eqref{10}, we can re-express $\mathbb{P}$ as
\begin{equation}
\mathbb{P}=-(\frac{53.4}{F^{2/3}})~M_{U}^{2}\big( -\frac{1}{s}(.690F^{4/3})\beta l + \frac{3l}{t} \big) \ .
\label{70}
\end{equation}
Hence, it follows from \eqref{65} that
\begin{equation}
V_{D}=(\frac{1.42 \times 10^{3}}{F^{4/3}})\frac{M_{U}^{4}}{s}\big( -\frac{1}{s}(.690F^{4/3})\beta l + \frac{3l}{t} \big)^{2} \ .
\label{71}
\end{equation}
Note that $V_{D}$ is a function of $s$ and $t$ (the real parts of $S$ and $T$ respectively) and is independent of the ``axions'' $\sigma$ and $\chi$.

Finally, we note that the part of $\mathbb{P}$ which is independent of the charged matter scalars,  that is, $\mathbb{P}$ given in \eqref{69}, when evaluated for some fixed values of $\langle s \rangle$ and $\langle t \rangle$ is customarily referred to as the Fayet-Iliopoulos (FI) term. We will do so, henceforth, in this paper.

\subsection{Supersymmetry Preserving Vacua of $V_{D}$}

Minimizing the D-term potential \eqref{71} defines $D$-flat, $N=1$ supersymmetry preserving vacuum states $\langle s \rangle$,$\langle t \rangle$ for which
%
\begin{equation}
\langle \mathbb{P} \rangle=FI=0 \ .
\label{72}
\end{equation}
It follows from \eqref{69} that this will be the case for 
\begin{equation}
\langle s \rangle=\frac{\epsilon_{S}^{\prime}\beta}{3} \langle t \rangle
\label {73}
\end{equation}
or, using \eqref{64}, for
\begin{equation}
\langle s \rangle=.230 F^{4/3}\beta \langle t \rangle \ .
\label{74}
\end{equation}
Integer $l$ which defines the line bundle $L$ has canceled out of this expression. Importantly, note that the values of $\langle s\rangle$,$\langle t \rangle$ are {\it not completely determined} by requiring that $FI=0$. Rather, they form a straight one-dimensional line in $s$ and $t$ space where $V_{D}=0$. That is, {\it a priori, $\langle t \rangle$ can take any value whereas $\langle s \rangle$ is constrained to satisfy \eqref{74}}. The value for $\langle t \rangle$ will be fixed to be a minimum of the F-term potential $V_{F}$ in Section 6.  Finally, note that since $V_{D}$ is independent of the axions $\sigma$ and $\chi$, there is, as yet, no constraints on their vacuum expectation values.

Let us choose {\it any} point along the $D$-flat direction and expand the complex fields $S$ and $T$ around the associated vacuum expectation values. Expanding
\begin{equation}
S=\langle s \rangle + \delta S~~,~~T=\langle t \rangle + \delta T
\label{75}
\end{equation}
one finds that the Lagrangian for $\delta S$ and $\delta T$ has off-diagonal kinetic energy and mass terms. However, as shown in \cite{Dumitru:2022apw}, one can define two new complex fields $\xi^{1}$ 
and $\xi^{2}$ which have canonically normalized kinetic energy and are mass eigenstates.

Following \cite{Dumitru:2022apw}, we define
\begin{equation}
\begin{pmatrix} 
\xi^{1} \\
\xi^{2} 
\end{pmatrix}
=U
\begin{pmatrix} 
\delta S \\
\delta T
\end{pmatrix}
\label{76}
\end{equation}
where
\begin{equation}
U=\frac{1}{\langle \Sigma \rangle}
\begin{pmatrix}
\langle g_{S\bar{S}} \bar{k}_{S}\rangle & \langle g_{T\bar{T}} \bar{k}_{T}\rangle \\
\\
\sqrt{\langle g_{S\bar{S}}g_{T\bar{T}} \rangle}\langle  \bar{k}_{T}\rangle & -\sqrt{\langle g_{S\bar{S}}g_{T\bar{T}} \rangle}\langle  \bar{k}_{S}\rangle
\end{pmatrix}
\label{77}
\end{equation}
and
\begin{equation}
\Sigma^{2}= g_{S \bar{S}}k_{S}\bar{k}_{S}+ g_{T \bar{T}}k_{T}\bar{k}_{T} \ .
\label{78}
\end{equation}
It then follows that the Lagrangian for $\xi^{1}$ and $\xi^{2}$ is given by
\begin{equation}
\mathcal{L}=-\partial^{\mu}\bar{\xi}^{1}\partial_{\mu}\xi^{1}-\partial^{\mu}\bar{\xi}^{2}\partial_{\mu}\xi^{2} - m_{\rm anom}^{2}\bar{\xi}^{1}\xi^{1} \ ,
\label{79}
\end{equation}
where 
\begin{equation}
m_{\rm anom}=\sqrt{2\langle g_{2}^{2}\Sigma^{2}\rangle} 
\label{80}
\end{equation}
and $\xi^{2}$ is massless. As discussed below, $m_{\rm anom}$ is the mass of the gauge connection of the anomalous $U(1)$ structure group.

Let us evaluate $m_{\rm anom}$. The gauge coupling is defined by
\begin{equation}
g_{2}^{2}=\frac{\pi \hat{\alpha}_{GUT}}{s}  \ .
\label{81}
\end{equation}
Using\eqref{38} and \eqref{74}, we find that in the chosen $FI=0$ vacuum
\begin{equation}
\langle g_{2} \rangle=\frac{.489F^{1/3}}{\langle s \rangle^{1/2}} =(\frac{1.02}{F^{1/3}\beta^{1/2}}) \frac{1}{\langle t \rangle^{1/2}}\ .
\label{82}
\end{equation}
Computing $\Sigma^{2}$ in \eqref{78} using \eqref{25},\eqref{63} and \eqref{68} we find that
\begin{equation}
\Sigma^{2}=\frac{M_{P}^{2}}{8\pi}l^2\big(\frac{3.68 \times 10^{-5} F^{4/3}\beta^{2}}{s^{2}} +\frac{2.32 \times 10^{-4}}{t^{2}~F^{4/3}} \big) \ .
\label{83}
\end{equation}
Now evaluate this for the chosen $FI=0$ vacua. Using \eqref{74}, we find that the coefficient $\beta$ exactly cancels out of the expression. Finally, using \eqref{2a} and \eqref{10}, we find
\begin{equation}
\langle \Sigma \rangle=(\frac{2.35 \, l}{F^{2/3}})\frac{M_{U}}{\langle t \rangle} \ .
\label{84}
\end{equation}
It then follows from \eqref{80}, \eqref{82}  and \eqref{84} that
\begin{equation}
m_{\rm anom}=(\frac{3.39 \,l}{F\beta^{1/2}} )\frac{M_{U}}{\langle t \rangle^{3/2}} \ .
\label{85}
\end{equation}
Note that $m_{\rm anom}$ is a function of the expectation value $\langle t \rangle$.

Since complex fields $\xi^{1}$ and $\xi^{2}$ are mass eigenstates with canonical kinetic energy, it is useful to express both $S$ and $T$ in terms of $\xi^{1}$ and $\xi^{2}$. This can be done by inverting expression \eqref{76}. Then
\begin{equation}
\begin{pmatrix}
\delta S \\
\delta T
\end{pmatrix}
= U^{-1}
\begin{pmatrix}
\xi^{1} \\
\xi^{2}
\end{pmatrix}
\label{86}
\end{equation}
where
\begin{equation}
U^{-1}= \frac{1}{\langle \Sigma \rangle}
\begin{pmatrix}
\langle k_S \rangle &  \sqrt{\langle \frac{g_{T\bar{T}}}{g_{S\bar{S}} }\rangle} \langle k_{T} \rangle \\
\langle k_T \rangle &  -\sqrt{\langle \frac{g_{S\bar{S}}}{g_{T\bar{T}} }\rangle} \langle k_{S} \rangle
\end{pmatrix}
\label{87}
\end{equation}
Using \eqref{25},\eqref{63},\eqref{64},\eqref{68},\eqref{84}, as well as \eqref{2a} and \eqref{10}, we find that
\\
\begin{equation}
U^{-1}= \frac{i \langle t \rangle}{M_{P}}
\begin{pmatrix}
2.00 F^{4/3} \beta & -1.15 F^{4/3}\beta \\
\\
-2.90 & -5.00
\end{pmatrix}
\label{88}
\end{equation}
Writing
\begin{equation}
\delta S= \delta s+i\sigma~~,~~\delta T= \delta t+i\chi \ ,
\label{89}
\end{equation}
it follows from \eqref{86} and \eqref{88} that
\begin{equation}
\begin{aligned}
&\delta s+i\sigma=\frac{i \langle t \rangle}{M_{P}}\big(2.00 F^{4/3} \beta ~\xi^{1} - 1.15 F^{4/3}\beta~\xi^{2} \big), \\
&\delta t +i\chi= \frac{i \langle t \rangle}{M_{P}}\big(-2.90~\xi^{1} - 5.00~\xi^{2} \big) \ .
\label{90}
\end{aligned}
\end{equation}

To proceed, we recall that $\xi^{1}$ is a massive complex field with mass $m_{\rm anom}$ given in \eqref{85}, while $\xi^{2}$ is massless. Defining
\begin{equation}
\xi^{1}=\eta^{1}+i\phi^{1} \ ,
\label{91a}
\end{equation}
it was shown in detail in \cite{Ashmore:2020ocb, Dumitru:2021jlh, Dumitru:2022apw} that the anomalous $U(1)$ hidden sector gauge group is spontaneously broken such that: (1)   the $U(1)$ gauge field $A_{U(1)}$ attains a mass $m_{\rm anom}$; (2) the real scalar $\phi^{1}$ obtains the same mass $m_{\rm anom}$ based on  \eqref{79}; and, (3) real scalar $\eta^{1}$ acts as the zero mass Goldstone boson associated with this spontaneous breaking. In combination with the fermionic superpartner $\psi^{1}$ which also, by supersymmetry, must obtain mass $m_{\rm anom}$, the combination ($\phi^{1}$, $A_{U(1)}$, $\psi^{1}$) forms a vector supermultiplet with mass $m_{\rm anom}$. As we will demonstrate in a concrete example presented below, physically realistic values of $F$, $\beta$ , $l$ and $\langle t \rangle$ are typically such that 
\begin{equation}
m_{\rm anom} \gtrsim M_{U} 
\label{91}
\end{equation}
It follows from \eqref{9} that $M_{U}$ sets the Calabi-Yau threefold mass scale. In this paper, we are interested only in the effective low-energy theory composed of fields with mass substantially less than this scale, Hence, we will integrate out all fields with mass greater than or approximately equal to $M_{U}$. Thus, it follows from \eqref{91}, that we will integrate the anomalous $U(1)$ massive vector superfield, which includes the scalar field component $\phi^{1}$, out of the theory. Recalling that the Goldstone boson $\eta^{1}$ can be gauged away, it follows that the entire complex scalar field $\xi^{1}$ can be integrated out of the low energy theory.

Therefore, in the low energy effective theory, the expressions in \eqref{90} simplify to 
\begin{equation}
\begin{aligned}
&\delta s+i\sigma=-\frac{i \langle t \rangle}{M_{P}}\big(1.15 F^{4/3}\beta~\xi^{2} \big), \\
&\delta t +i\chi= -\frac{i \langle t \rangle}{M_{P}}\big(5.00~\xi^{2} \big) \ .
\label{92}
\end{aligned}
\end{equation}
Writing
\begin{equation}
\xi^{2}=\eta+i\phi \ ,
\label{93}
\end{equation}
it follows that
\begin{equation}
\begin{aligned}
&\delta s = \frac{\langle t \rangle}{M_{P}}1.15 F^{4/3}\beta~\phi~~,~~\sigma=-\frac{\langle t \rangle}{M_{P}} 1.15 F^{4/3}\beta~ \eta \\
&\delta t= \frac{\langle t \rangle}{M_{P}}5.00~\phi~~,~~\chi=-\frac{\langle t \rangle}{M_{P}}5.00~ \eta \ .
\end{aligned}
\label{94}
\end{equation}
It then follows from \eqref{75}, \eqref{89} and \eqref{94} that
\begin{equation}
\begin{aligned}
&s= \langle s \rangle+\frac{\langle t \rangle}{M_{P}}1.12F^{4/3}\beta~\phi~~,~~t=  \langle t \rangle+\frac{\langle t \rangle}{M_{P}}5.00~\phi \\
&\sigma=-\frac{\langle t \rangle}{M_{P}} 1.15 F^{4/3}\beta~ \eta~~,~~\chi=-\frac{\langle t \rangle}{M_{P}}5.00~ \eta \ .
\end{aligned}
\label{95}
\end{equation}
Now recall from \eqref{74} that
\begin{equation}
\langle s \rangle=.230 F^{4/3}\beta \langle t \rangle \ .
\label{96}
\end{equation}
Inserting this into \eqref{95} we obtain finally that
\begin{equation}
\begin{aligned}
&s= \langle t  \rangle F^{4/3}\beta (.230 +1.15~\frac{\phi}{M_{P}})~~,~~t=  \langle t \rangle(1+ 5.00~\frac{\phi}{M_{P}}) \\
&\sigma=-\langle t \rangle 1.15 F^{4/3}\beta~ \frac{\eta}{M_{P}}~~,~~\chi=-\langle t \rangle5.00~\frac{\eta}{M_{P}} \ .
\end{aligned}
\label{97}
\end{equation}
It is {\it important} to note that $s$ and $t$ in this expression satisfy
\begin{equation}
s=.230 F^{4/3}\beta~t \ ,
\label{98}
\end{equation}
that is, the same relationship as in \eqref{76}. Hence, the expansion of $s$ and $t$ around vacuum expectation values $\langle s \rangle$ and $\langle t \rangle$ satisfying $FI=0$, and integrating out the heavy scalar $\xi^{1}$ while keeping the light scalar $\xi^{2}$ in the effective theory, restricts all values of $s$ and $t$ to lie along the $D$-flat direction 
\begin{equation}
s =\frac{\epsilon_{S}^{\prime}\beta}{3} t = .230 F^{4/3}\beta~t
\label {99}
\end{equation}
in which $V_{D}=0$.

\section{Determining the Moduli Vacua}

In the previous Section, we constructed the D-term potential energy $V_{D}$ for the dilaton and K\"ahler modulus induced by the their inhomogeneous transformations under the anomalous $U(1)$ gauge group. The minima of this D-term potential energy was shown to lie along a one-dimensional line in $s$ and $t$ space for which $V_{D}=0$. Specifically, $FI=0$ and, hence, $N=1$ supersymmetry is preserved for $s$ and $t$ satisfying relation \eqref{99}.
 
 In this Section, we will explicitly assume that parameters $F$, $\beta$, $l$ and the values of $\langle t \rangle$ are restricted so that $m_{\rm anom} \gtrsim M_{U}$.
%
%
It then follows from \eqref{85} that, for fixed values of $F$, $\beta$ and $l$, there is a maximum value of $\langle t \rangle$, which we denote $\langle t \rangle_{bound}$, given by 
\begin{equation}
\langle t \rangle_{bound}= \large( \frac{3.39 l}{F\beta^{1/2}} \large)^{2/3}  \ .
\label{new1}
\end{equation}
Below this bound, one can integrate complex scalar $\xi^{1}$ out of the low energy theory, leaving $s$, $t$, $\sigma$ and $\chi$ to be expressed in 
terms of $\langle t \rangle$, $\phi$ and $\eta$ as in \eqref{97}. In this Section, we analyze the F-term potential given in \eqref{58} along the $V_{D}=0$ line for $\langle t \rangle \lesssim \langle t \rangle_{bound}$ using \eqref{97}. We will discuss the case where $\langle t \rangle \gg \langle t \rangle_{bound}$ in the next Section. 

Finally, for concreteness, we henceforth assume that the commutant subgroup to the anomalous $U(1)$ is $E_{7}$ and, therefore, that $b_{L}=6$ -- as stated in \eqref{39}. Using the expression for $\hat{\alpha_{GUT}}$ in \eqref{38}, it follows that the parameter $b$ in \eqref{37} is given by
\begin{equation}
b=\frac{41.2}{F^{2/3}} \ .
\label{99a}
\end{equation}

\subsection{The F-term Potential Along the $V_{D}=0$ Line}
Inserting expressions \eqref{97} for $s$, $t$, $\sigma$ and $\chi$ into the F-term potential energy $V_{F}$ given in \eqref{58}, and using $b$ given in \eqref{99a}, we find 
\begin{equation}
\begin{aligned}
V_{F} (\langle t \rangle, \tilde{\eta}, \tilde{\phi}) = &\frac{M_{U}^{4}}{F^{4/3} \beta \langle t \rangle^{4}\langle c \rangle^{3} (0.230 +1.15 \tilde{\phi}) (1+5.00 \tilde{\phi})^3}  \\ 
 \times & \left[ 1.138 \, F^{-4/3} \tilde{d} (A^2+B^2)  \right. \\ 
+& 1.32 \times 10^{-6} \tilde{d}^{-1}\big((1+19.0 F^{2/3}\beta \langle t \rangle(1+5.01 \tilde{\phi})^{2}+3 \big) \\ &  \times \exp[-19.0\,  F^{2/3} \beta \langle t \rangle (1+5.01 \,  \tilde{\phi})]  \\
-&(2.43 \times 10^{-3} F^{-2/3} ) \, \big(1+ 19.0 \, F^{2/3} \beta \langle t \rangle (1+5.01\, \tilde{\phi})    \big) \\ & \times
\exp[-  9.48 \,  F^{2/3} \beta \langle t \rangle (1+5.00 \, \tilde{\phi}) ] \\ & \times sgn(A)\sqrt{A^2+B^2} 
\cos[47.5 \,  F^{2/3} \beta \langle t \rangle \tilde{\eta} -\arctan(\frac{B}{A})] \\
+&2.62 \times 10^{-6} \tilde{d}^{-1}\,  p  \, \big(5.50+ \langle t \rangle(19.0 \,  F^{2/3} \beta (1+5.01 \, \tilde{\phi}) + 3 \tau (1+5.00 \,  \tilde{\phi})    )\big) \\ & \times
\exp[ -(9.49 \, F^{2/3} \beta (1+5.01 \, \tilde{\phi}) + \tau (1+5.00 \, \tilde{\phi}))\langle t \rangle ]  \\ & \times \cos[(-47.5 \,  F^{2/3} \beta + 5.00 \,  \tau) \langle t \rangle \tilde{\eta} +\theta_{p}]\\
+&4.36 \times 10^{-7} \tilde{d}^{-1}p^{2} \big(3+(3+2 \, \tau \langle t \rangle (1+ 5.005 \, \tilde{\phi}))^{2} \big)
\\ & \times \exp[-2\tau \langle t \rangle (1+ 5.00 \, \tilde{\phi})]  \\
-& 2.43  \times 10^{-3} \, F^{-2/3}  \,  p \, (1+2\tau \langle t \rangle (1+5.00  \, \tilde{\phi}))
\\ &  \times \exp[-\tau \langle t \rangle (1+ 5.00 \,  \tilde{\phi})] \\ & 
\left.  \times sgn(A)\sqrt{A^2+B^2} \cos[5.00 \, \tau  \langle t \rangle \tilde{\eta}+\theta_{p} - \arctan(\frac{B}{A})]  \right] 
\end{aligned}
\label{PC1}
\end{equation}
where $\tilde{\eta}= \eta/M_P$ and $\tilde{\phi}= \phi/M_P$ are dimensionless. As in \eqref{58}, coefficients $A$, $B$, $\langle c \rangle$ are defined in \eqref{23b}, $F$ is defined in \eqref{11},\eqref{12}, $\tau$ is given in \eqref{51} and $p,\theta_{p}$ arise in \eqref{55}.

This is a rather complicated expression which is difficult to analyze analytically. We find it most illuminating to simply choose some fixed values for the flux parameters $A$, $B$, $\langle c \rangle$, and the coefficients $F$, $\beta$ subject to the constraints on them discussed above. We will also choose a realistic fixed value for $\tau$ and set $\tilde{d}=1$. The potential $V_{F}$ will then be evaluated at these fixed parameters, initially allowing parameters $p$ and $\theta_{p}$ to take arbitrary values. The choices of $F$, $\beta$ and $l$ will also, using \eqref{new1}, determine a fixed value for $\langle t \rangle_{bound}$. We then plot $V_{F}$ in the range $0 \leq \langle t \rangle \lesssim \langle t \rangle_{bound}$ for various values of $p$ and $\theta_{p}$. 

\subsection{Stabilizing the Axion}

It is important to note that the ``light'' axion $\tilde{\eta}$ enters $V_{F}$ in \eqref{PC1} via three cosine terms, specifically in the third, fourth and sixth terms, each with a different coefficient that depends on the parameters listed above as well as on $\langle t \rangle$. Therefore, before discussing the stabilization of $\langle t \rangle$, it is essential to determine the vacuum expectation value of $\tilde{\eta}$. To do this, we first note that the coefficient of each of the three cosines has an exponentially suppressed multiplicative factor. For the third, fourth and sixth terms in \eqref{PC1} -- that is, the terms containing the cosines -- these factors are
\begin{equation}
\exp[-  9.49 \,  F^{2/3} \beta \langle t \rangle]~,~\exp[ -(9.49 \, F^{2/3} \beta + \tau )\langle t \rangle ]~,~ \exp[-\tau \langle t \rangle ] 
\label{PC1a}
\end{equation}
respectively, where we have ignored the $\tilde{\phi}$ field contribution which simply enhances the suppression of each term equally.  For the physically acceptable range of parameters $F$, $\beta$ and $\tau$ discussed above, we find that the value of $\langle t \rangle$ is such that the first two exponentials in \eqref{PC1a} are {\it greatly} suppressed relative to the third entry -- typically by a factor of $10^{-6}$ or smaller. Hence, a {\it very} good approximation for determining $\langle \tilde{\eta} \rangle$ is to drop the third and fourth terms in \eqref{PC1} and to evaluate $\langle \tilde{\eta} \rangle$ using the sixth term only. Then, using the fact that there is a minus sign in front of the sixth term, we find
\begin{equation}
\frac{\partial V_{F}}{\partial\tilde{\eta}} =0,~~~\frac{\partial^{2}V_{F}}{\partial^{2}{\tilde{\eta}} } >  0~~ \Rightarrow ~~\langle \tilde{\eta} \rangle = \frac{2\pi n+{\rm arctan}(\frac{B}{A}) -\theta_{p}} {5.005 \tau \langle t \rangle} 
\label{PC1b}
\end{equation}
where $n \in \mathbb{Z} $. We conclude that once we have found the local minimum for $\langle t \rangle$ in the next subsection, substituting its value into \eqref{PC1b} gives one a very good approximation for the local minima of $\tilde{\eta}$ as a function of the Pfaffian parameter $\theta_{p}$. As a check, we calculated the minima for $\tilde{\eta}$ including all three cosine terms using Mathematica. We find that expression \eqref{PC1b} is indeed the correct expression for $\langle \tilde{\eta} \rangle$ to a very high degree of accuracy. That is, {\it for any value of $\theta_{p}$ }we have stabilized the ``light'' axion. Having done this, we now proceed to computing the vacuum expectation value for $\langle t \rangle$ which minimizes the $V_{F}$ potential energy.

\subsection{Stabilizing $\langle t \rangle$: An Explicit Example}

In this paper, we will present only a single, but physically representative, set of parameters as an example. A much more comprehensive discussion of $V_{F}$ will be presented elsewhere.

First, recall that $V_{F}$ is a function of $\langle t \rangle$, $\tilde{\phi}$ and $\tilde{\eta}$. It is clear from the \eqref{97} that $\tilde{\phi}$ is simply the fluctuation around $\langle t \rangle$ in the $t$ direction. Hence, to evaluate $\langle t \rangle$ one can simply set $\tilde{\phi}=0$. Furthermore, as stated above, to a very high degree of accuracy one can approximate $\tilde{\eta}$ by the vacuum expectation value given in \eqref{PC1b}--which sets the associated cosine factor to unity for {\it any} value of $\theta_{p}$. Having done this, $V_{F}$ becomes a function of $\langle t \rangle$ only. We emphasize that, although we have used \eqref{PC1b} as an approximation to $\langle \tilde{\eta} \rangle$, the following calculation uses {\it all terms} for $V_{F}$ given in \eqref{PC1}.
In Figure 1, we show a sequence of potentials $V_{F}$ corresponding to fixed flux coefficients
$A=2/3$, $B=A/\sqrt{3}$, $c=\frac{1}{\sqrt{3}}$, as well as choosing $\beta=2$, $l=2$, $F=2$ and $\tau=2$. However, we allow the Pfaffian coefficient $p$ to vary over a range of values.  
It follows from \eqref{new1} that,  for this choice of parameters,
%
%
 %
\begin{equation}
\langle t \rangle_{bound}=1.79 \ .
\label{B}
\end{equation}
Therefore, we plot these curves in the region where $0\leq\langle t  \rangle \leq 1.79$.
From bottom to top, the shapes range from having a global minimum with $V_F<0$; a global minimum with $V_F=0$;  a local minimum with $V_F>0$;  an inflection point; and no extremum at all.  We find that a similar range of potential shapes satisfying  these same approximation conditions can be obtained for $1 < \tau < 3.75$ (keeping all other parameters fixed).

\begin{figure*}[!h]
 \begin{center}
\includegraphics[width=5.5in,angle=-0]{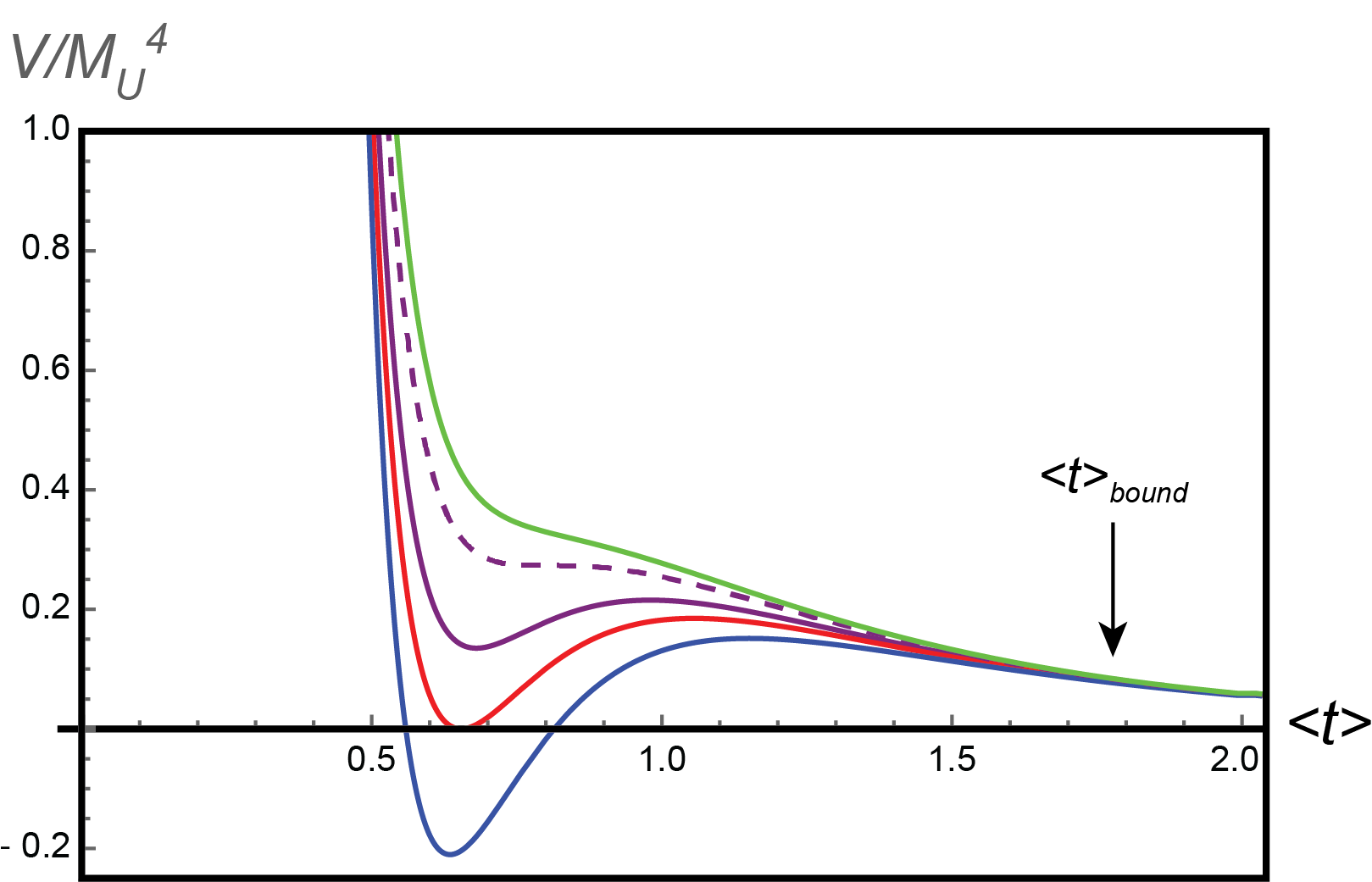}
\end{center}
\caption{ Range of potential shapes  for fixed $A=2/3$, $B=A/\sqrt{3}$, $c=\frac{1}{\sqrt{3}}$ as well as $l=2$, $\beta=2$, $F=2$, $\tau=2$ and an arbitrary parameter $\theta_{p}$,  but varying the Pfaffian coefficient $p$.  
 The values of $p$ for the curves from the bottom to the top are $p=(370,347,333,316,307)$ respectively. $\langle t \rangle_{bound}$ represents the value of $\langle t \rangle$  above which  $m_{\rm anom}<sM_{U}$ and so the potential shape may not be precise.}

\end{figure*} 

Whether there is a $V_{F}$ extremum at a given $\langle t \rangle$ depends on the value of $p$. Choosing an explicit value of $p$, one can attempt to solve $dV_{F}/d\langle t \rangle=0$ over the range  $0 \leq \langle t \rangle \leq \langle t \rangle_{bound}$. For some choices of $p$, there will be no solution and, hence, there are no extrema of the potential $V_{F}$ over the 
allowed range of t. This is the case for the green curve in Figure 1.
For a special choice of $p$, there will be a solution of $dV_{F}/d\langle t \rangle=0$ at a single value of $\langle t \rangle$ in the given range. This leads to an inflection point in the potential energy. This is the case for the dashed purple curve in Figure 1.  For a finite range of values of $p$, we find solutions with two extrema, one corresponding to a minimum and the other to a local maximum of the potential energy. This is the case for the blue, red and solid purple curves in Figure 1. The values of $p$ leading to each of these results are given in the caption for Figure 1.
Note that the value of $p$ is progressively decreasing from the bottom curve (blue, with a negative energy global minimum) through the red curve (zero energy global minimum), solid purple curve (positive energy local minimum), dashed purple curve (positive energy inflection point) and green curve (no extremum). Experimenting over a substantial range of parameters $p$, we find that a large number give a potential energy with a negative global minimum while a smaller, but substantial, number lead to a zero or positive potential energy minimum. Finally, as a check on these results, we performed the above calculations using all three cosines to determine $\langle \tilde{\eta} \rangle$, rather than simply inserting \eqref{PC1b}. The results are identical to the above to at least decimal places.

Given potential energy \eqref{PC1}, one can determine the masses of the dimensionful fluctuation fields $\phi ~(=\tilde{\phi}M_{P})$ and $\eta ~(=\tilde{\eta}M_{P})$ by computing $\partial^2 V_F / \partial^2 \phi$ and $\partial^2 V_F / \partial^2 \eta$ evaluated at the minima at $\langle t \rangle <  \langle t \rangle_{bound}$ for the blue, red and solid purple curves, respectively. These masses, along with the associated value of $m_{\rm anom}$ computed at each such $\langle t \rangle$ using \eqref{85}, are presented in Table 1.
\begin{table}
\begin{center}
\begin{tabular}{ |c | c | c | c|}
        \hline
        & $m_{\rm anom}$ & $m_{\phi}$ & $m_{\eta}$ \\
        \hline
 \; blue ($V_{\rm min} <0$)   \;     & \;  $1.5 \times 10^{17}$~ GeV  \; & \;  $1.6 \times 10^{15}$~ GeV  \; & \;  $1.2 \times 10^{15}$~ GeV \;  \\
  \hline
 \;   red ($V_{\rm min}=0$)    \;    & \;  $1.4 \times 10^{17}$~ GeV  \; &  \; $1.3 \times 10^{15}$~ GeV \;  & $ \; 1.7 \times 10^{15}$~ GeV \;  \\
    \hline
 \; solid purple ($V_{\rm min} > 0$)    \;    &  \; $1.0 \times 10^{17}$~ GeV  \; & \;  $1.1 \times 10^{15}$~ GeV  \;  & \;  $1.4 \times 10^{15}$~ GeV \;  \\
         \hline
        \end{tabular}
        \caption{The values for $m_{\rm anom}$ and $m_{\phi}$, $m_{\eta}$ at the minima of the blue, red and solid purple curves in Figure 1. }
          \end{center}
  \end{table}
Note that all values of $m_{\rm anom}$ exceed $M_{U}$, as they must. Importantly, in all three cases the values of the $m_{\phi}$ and $m_{\eta}$ are each over an order of magnitude smaller than $M_{U}=3.15 \times 10^{16} {\rm GeV}$. Hence, the $\phi$ and $\eta$ moduli remain in the low energy effective field theory.

\subsection{Finding the Range of the Pfaffian Coefficient $p$}

Note that the coefficient $F=2$ chosen for $V_{F}$ associated with Figure 1, is the largest value of $F$ in the  ``standard'' range given in \eqref{12}. However, as alluded to in Subsection 2.2, it is possible in specific vacua for the value of the coefficient $F$ to either exceed or be less than the upper and lower bounds respectively presented in \eqref{12}.
The vacua discussed in this paper are exactly of this type. The reason is the following. It has been shown \cite{Lukas:1998tt, ovrut2020vacuum} that in generic heterotic M-theory vacua, the ``effective'' expansion parameter is given by
\begin{equation}
\epsilon_{S}^{\rm eff}= 2\epsilon_{S}^{\prime}\frac{t}{s} \ ,
\label{paris1}
\end{equation}
where we have used the fact that $s=V$ and that, in the $h^{1,1}=1$ case, $\hat{R}=2t$. It then follows from \eqref{99} that
\begin{equation}
\epsilon_{S}^{\rm eff}= \frac{6}{\beta} \ .
\label{paris2}
\end{equation}
We conclude that in the vacua we are considering -- that is, the $h^{1,1} =1$ case where $s$ and $t$ satisfy $FI=0$, thus setting $V_{D}=0$ -- the size of the effective strong coupling parameter is set by parameter $\beta$, but is {\it independent} of the coefficient  $F$. Hence, $F$ need not be bounded by the constraints in \eqref{12} and can, {\it in principle}, take any value -- including values larger than $2$ and smaller than $0.6$. 

However, in considering possible shapes of the potential energy curve $V_F$, the value of $F$ cannot be made arbitrarily large. This is because  the condition  $m_{anom} \ge M_U$ has been assumed in deriving $V_F$, and this can only be satisfied for $0 \leq \langle t \rangle \lesssim \langle t \rangle_{bound}$ where, from \eqref{new1},  $ \langle t \rangle_{bound} \propto 1/F^{2/3}$.  To be sure of  potentials with stable or metastable minima and local maxima, like the bottom three curves in Figure~1, it must be that $ \langle t \rangle_{bound}$ exceeds $\langle t \rangle$ at the local maximum for each curve, which sets an upper bound for $F$. 
As a concrete example, let us consider the blue curve in Figure 1. This curve was determined using $F=2$, leading to $\langle t \rangle_{bound} =1.79$ given in \eqref{B}. It will be shown in a subsequent publication \cite{Paper2} that the values of $\langle t \rangle$ at both the minima and maxima of any $V_{F}$ curve will remain unchanged under a change in parameter $F$, as long as the Pfaffian parameter is appropriately adjusted. 

Let us denote the value of $\langle t \rangle$ at the local maximum of the blue curve by $\langle t \rangle_{max}$. We see from Figure 1 that $\langle t \rangle_{max}=1.1$. Let us now gradually raise the value of $F$ to lower the  value of $\langle t \rangle_{bound}$  -- but appropriately adjusting $p$ at each stage so that the extrema of the blue curve remain at the same values of $\langle t \rangle$.  We can continue until we reach the point where
\begin{equation}
\langle t \rangle_{bound}=\langle t \rangle_{max}=1.1  \ .
\label{yellow}
\end{equation}
This occurs when $F$ reaches $F=4$, in accordance with \eqref{new1}  (and using the values of $l $ and $\beta$ cited in the caption of  Figure 1).   Note that this is significantly larger than the conventional upper bound $F=2$ given in \eqref{12} and used in constructing Figure 1.  As will be  shown in \cite{Paper2}, the appropriate adjustment of $p$ is to rescale, $p \propto F^{-2/3}$.  
 In other words, in changing from $F=2$  to $F=4$, the values of $ \langle t \rangle$ at the minima and maxima do not change if one adjusts the corresponding value of Pfaffian parameter  substantially from $p=370$ to $p=233$. 
 A similar adjustment  is required for the two higher  curves--that is, the red and solid purple curves--in Figure 1.  The values of $\langle t \rangle$ at their maxima, the values of $F_{max}$ and the associated values of $p$ for each of these three curves are given in Table 2.
\begin{table}
\begin{center}
\begin{tabular}{ |c | c | c | c|}
      \hline
        & $\langle t \rangle$ at max   & $F_{max}$ & $p$  \\
        \hline
 \; blue ($V_{\rm min} <0$)     \;   & \;  1.1  \; & \;  4.0  \; &  \; 233 \; \\
  \hline
 \;   red ($V_{\rm min}=0$)   \;     & \;  1.0 \;  & \; 4.4  \; & \;  205 \;  \\
    \hline
 \; solid ($V_{\rm min} > 0$)      \;  & \;  0.97 \;  &  \; 4.8 \;  & 185 \;  \\
         \hline
        \end{tabular}
        \caption{The values of $\langle t \rangle$ at the local maximum and the associated values for $F_{max}$ and $p$ when $\langle t  \rangle _{bound}$ is set equal to $\langle t \rangle$ maximum, for the blue, red and solid purple curves in Figure 1 respectively.}
  \end{center}
 \end{table}
Having done this, it is important to check that the masses of the fields $\phi$ and $\eta$ at the local minima of each of these three curves continue to be considerably smaller than the value of $m_{anom}$ evaluated at each such local minimum. This turns out to to be the case, as is shown explicitly in Table 3.
 \begin{table}
\begin{center}
\begin{tabular}{ |c | c | c | c|}
        \hline
        & $m_{\rm anom}$ & $m_{\phi}$ & $m_{\eta}$ \\
        \hline
 \; blue ($V_{\rm min} <0$)    \;    & \;  $7.1 \times 10^{16}$~ GeV  \; &  \; $6.5 \times 10^{14}$~ GeV \;  & \;  $4.7 \times 10^{14}$~ GeV  \; \\
  \hline
  \;  red ($V_{\rm min}=0$)    \;    &  \; $6.5 \times 10^{16}$~ GeV  \; & \;  $5.8 \times 10^{14}$~ GeV \;  & \;  $4.2 \times 10^{14}$~ GeV \;  \\
    \hline
 \; solid purple ($V_{\rm min} > 0$)     \;   &  \; $5.9\times 10^{16}$~ GeV  \; & \;  $5.1 \times 10^{14}$~ GeV \;  & \;  $3.7 \times 10^{14}$~ GeV \;  \\
         \hline
        \end{tabular}
        \caption{The values of $m_{anom}$ and $m_{\phi}$, $m_{\eta}$  for each of the three curves defined in Table 2. Note that, in all cases, $m_{\phi}$ and $m_{\eta}$ are each $\ll m_{anom}$. }
  \end{center}
  \end{table}

We conclude that by lowering $\langle t \rangle_{bound}$ to the values of the local maxima for each of the blue, red and solid purple curves in Figure 1,  the values of the Pfaffian coefficient $p$ vary through a substantial range. As a second example, a red curve with vanishing potential energy at the same value of $\langle t \rangle$ is possible for the Pfaffian coefficient range $205 <  p < 347$. 
The ability to stabilize the expectation values of dilaton and geometric moduli at the same value for a wide range of the Pfaffian coefficient $p$ has an important implication. As mentioned in Section 4, 
when fixed at a supersymmetry preserving minimum of the vector bundle moduli, the Pfaffian becomes a complex number specified in \eqref{home1}  by  an amplitude $p$ and a phase $\theta_{p}$.
As discussed above, it is not known how to explicitly calculate their values. However, it is trivial to show that the values of $\langle s \rangle$ and $\langle t \rangle$ at the potential minimum do not depend on $\theta_p$ at all; and now, as we have demonstrated, their values can also  be obtained for a wide range of $p$.  
The fact that the dilaton and geometric moduli can have the same expectation values  for a wide range of Pfaffian parameter $p$ means there is more likely to be some vector bundle moduli vacuum that produces a $p$ in that range and stabilizes those values. 
The explicit method for determining  the Pfaffian parameter $p$ for each type of potential in Figure~1 will be presented in detail in a subsequent publication \cite{Paper2}.


\section{Swampland Bound on the Potential at Large Values of $t$}

Our explicit construction of potentials  for heterotic $M$-theory compactified on Calabi-Yau threefolds with $h^{1,1}=h^{2,1}=1$ provides interesting test cases for the Swampland conjectures \cite{Ooguri:2006in, Ooguri:2018wrx, Lust:2019zwm, Palti:2019pca, Bedroya:2019snp,Rudelius:2021azq}.  Among the different  conjectures, the Transplanckian Censorship Conjecture \cite{Bedroya:2019snp} and the Strong de~Sitter Conjecture \cite{Rudelius:2021azq} 
both postulate that, for {\it large} values of the moduli fields (with canonically normalized kinetic energy), there is a positive lower bound on the gradient of the potential   when $V>0$, namely
\begin{equation}
\frac{|\nabla V|}{V} \ge \frac{2}{\sqrt{d-2}},
\label{burt1}
\end{equation}
where $d$ is the spacetime dimension.  
Since $d=4$ in our case,  the Swampland lower bound is equal to $\sqrt{2}$.

To evaluate whether our potential satisfies this Swampland conjecture we first recall that the total potential energy density is $V=V_{D}+V_{F}$, where $V_{D}$ is the function of $s$ and $t$  given in \eqref{71}. Since we are interested in the large $t$ limit where $t \gg t_{bound}$, $V_{F}$ is given in \eqref{58} (and {\it not} by \eqref{PC1}, which assumes $t \lesssim t_{bound}$). Now, as discussed in Subsection 5.3, $V_{D}=0$ for $s =.2301F^{4/3}\beta t$, which we assume henceforth. Inserting this into \eqref{58}, it follows that $V_{F}$ is a function of the modulus field $t$ and the axions $\sigma$ and $\chi$.  Although this $V_F$ has numerous terms, the task  of evaluating its large field limit is straightforward. Except for the first term, all other terms  in $V_{F}$ are suppressed by a factor of  ${\rm exp} (- {\bf c }t )$ for some positive coefficient ${\bf{c}}$ (which differs for each of the terms). This includes  all three terms containing the axions $\sigma$ and $\chi$,   so it is not necessary to consider their large field  limits. Hence, the first term dominates all other terms in the large $t$ limit.  Keeping only the first term in $V_{F}$, we have 
 \begin{equation}
 V_F \propto \frac{1}{s t^3} \propto \frac{1}{t^4}
 \label{burt2}
 \end{equation}
 where have imposed the condition that $s \propto t$ along the $D$-flat direction.  We do not need to use the exact proportionality constant because the Swampland condition uses the logarithmic derivative of $V_F$, so any constant factors drop out.

 To proceed, we note that the Swampland condition \eqref{burt1} requires the moduli fields to have canonically normalized kinetic energy. 
 The kinetic energy for  $t = {\rm Re} \, T$  is given by
\begin{equation}
\kappa^{2}_{4}  \frac{\partial^2 K}{ \partial T\partial \bar{T} }  ( \partial T  \partial \bar{T})|_{{\rm Im }T =0}  = \frac{3}{4} \frac{(\partial t)^2}{t^2}
\label{KE1}
\end{equation}
where  $\kappa^{2}_{4}K = -3 \ln \, (T+\bar{T} )$ is the K\"{a}hler potential given in \eqref{25}.
To rewrite this kinetic energy in terms of a field $\Phi$ with canonical kinetic energy, we use the
ansatz  $\Phi \equiv q \, \ln t$, where $q$ is a constant to be determined by the condition that  $\Phi$ has canonical kinetic energy.  That is, we set 
\begin{equation}
\frac{1}{2} (\partial \Phi)^2 = \frac{q^2}{2}  \frac{(\partial t)^2}{t^2}.
\label{KE2}
\end{equation}
equal to Eqs.~(\ref{KE1}) and obtain:
\begin{equation}
\frac{q^2}{2}= \frac{3}{4} ~~\Rightarrow~~q=\sqrt{3/2}.
\end{equation}
Hence,
 \begin{equation}
\Phi= \sqrt{3/2} \ln \, t \ .
\label{burt3}
\end{equation}

What remains, then, is 
 to rewrite $V_F$ in terms of canonical field $\Phi$ in the limit of large $t$ and to check if the Swampland constraint \eqref{burt1} is satisfied.  
Using \eqref{burt3}, we can rewrite the potential in \eqref{burt2} as
 \begin{equation}
 V_F \propto e^ {-4 \sqrt{2/3} \Phi},
 \label{burt4}
 \end{equation} 
from which we obtain
 \begin{equation}
\frac{|\nabla V|}{V} =\frac{|d V_F/d \Phi|}{V_F }= 4 \sqrt{2/3} > \sqrt{2} \ .
\label{ourlimit} 
 \end{equation}
 That is, our theory exceeds the Swampland lower bound of $\sqrt{2}$ in the large field limit \cite{Bedroya:2019snp}. A discussion of how our results relate to Swampland conjectures concerning conditions at small values of the $\langle t \rangle$ field near the center of moduli space 
 will be presented elsewhere \cite{Paper2}.
 
 \subsubsection*{Acknowledgements}

We thank Alek Bedroya and Cumrun Vafa for useful comments and for reviewing the manuscript.  Burt Ovrut is supported in part by both the research grant DOE No.~DESC0007901 and SAS Account 020-0188-2-010202-6603-0338. Ovrut would like to acknowledge the hospitality of the CCPP at New York University where much of this work was carried out.  Paul Steinhardt supported in part by the DOE grant number DEFG02-91ER40671 and by the Simons Foundation grant number 654561.   Steinhardt thanks Cumrun Vafa and the High Energy Physics group in the Department of Physics at Harvard University for graciously hosting him during his sabbatical leave.\\

\end{document}